\documentclass[aps,twocolumn,amsmath,amssymb,superscriptaddress]{revtex4-1}

\usepackage{graphicx}
\usepackage{dcolumn}
\usepackage{epsfig}
\usepackage{color}
\usepackage{ulem}
\usepackage{bbm}
\usepackage{amsmath}

\usepackage{hyperref,cleveref}

\def\beq{\begin{eqnarray}}
\def\eeq{\end{eqnarray}}
\def\beqa{\begin{eqnarray}}
\def\eeqa{\end{eqnarray}}

\newcommand{\be}{\begin{equation}}
\newcommand{\ee}{\end{equation}}

\def\bea{\begin{eqnarray}}
\def\eea{\end{eqnarray}}

\begin{document}

\title{Ferromagnetic fluctuations in the Rashba-Hubbard model}

\author{Andr\'es Greco}
\affiliation{Max-Planck-Institut f\"{u}r Festk\"{o}rperforschung, Heisenbergstra\ss{}e 1, D-70569 Stuttgart, Germany}
\affiliation{Facultad de Ciencias Exactas, Ingenier\'{\i}a y Agrimensura and Instituto de F\'{\i}sica Rosario (UNR-CONICET), Av. Pellegrini 250, 2000 Rosario, Argentina}

\author{Mat\'{\i}as Bejas}
\affiliation{Facultad de Ciencias Exactas, Ingenier\'{\i}a y Agrimensura and Instituto de F\'{\i}sica Rosario (UNR-CONICET), Av. Pellegrini 250, 2000 Rosario, Argentina}

\author{Andreas P.\ Schnyder}
\affiliation{Max-Planck-Institut f\"{u}r Festk\"{o}rperforschung, Heisenbergstra\ss{}e 1, D-70569 Stuttgart, Germany}

\date{\today}

\begin{abstract}
We study the occurrence and the origin of ferromagnetic fluctuations 
in the longitudinal spin susceptibility of the  $t$-$t'$-Rashba-Hubbard model on the square lattice.
The combined effect of the second-neighbor hopping $t'$ and the spin-orbit coupling leads to ferromagnetic fluctuations 
in a broad filling region. The spin-orbit coupling splits the energy bands, leading to two van Hove 
fillings, where the sheets of the Fermi surface change their topology. 
Between these two van Hove fillings the model shows ferromagnetic fluctuations. 
We find that the these ferromagnetic fluctuations originate from interband contributions to the spin
susceptibility. These interband contributions only arise if there is one holelike and
one electronlike Fermi surface, which is the case for fillings in between the two van Hove fillings.
 We discuss implications for experimental systems and propose a test on how
to identify these types of ferromagnetic fluctuations in experiments.
\end{abstract}

\maketitle

\section{Introduction}

Recent technological advances in atomic-scale synthesis have allowed to fabricate heterostructure interfaces with tailored electronic structures 
and symmetry properties~\cite{review_oxide_interfaces_Nature_12}. In these heterostructures it is possible, for example, 
to tune the degree of inversion-symmetry breaking and the strength of spin-orbit coupling by modulating  the layer thickness or by applying electric fields~\cite{matsuda_PRL_14,mizukami_matsuda_nat_phys_11,caviglia_triscone_Nature_08,sunko_king_nature_17,bhowal_npg_comp_mat_19,feng_levy_APL_15}. 
Many of these heterostructures exhibit emergent phenomena not
found in the bulk constituents~\cite{reyren_mannhart_science_07,brinkman_nat_mat_07,takahashi_tokura_APL_01,stemmer_PhysRevX_12,jackson_stemmer_PRB_13,gibert_triscone_nat_mat_12,suzuki_APL_mat_review_2015,nichols_lee_nat_commun_16,mazzola_king_PNAS_18}.  
Particularly interesting is the emergence of ferromagnetism at interfaces between correlated materials~\cite{brinkman_nat_mat_07,takahashi_tokura_APL_01,stemmer_PhysRevX_12,jackson_stemmer_PRB_13,gibert_triscone_nat_mat_12,suzuki_APL_mat_review_2015,nichols_lee_nat_commun_16,mazzola_king_PNAS_18},
as this could be of potential use for spintronics applications.
Interface ferromagnetism can arise both due to itinerant electrons, or due to localized spins at the interface.
The former case most likely occurs at surfaces of the delafossite oxides PdCoO$_2$ and PdCrO$_2$~\cite{mazzola_king_PNAS_18},
and at interfaces of GdTiO$_3$/SrTiO$_3$~\cite{stemmer_PhysRevX_12,jackson_stemmer_PRB_13}.
In order to understand how interface ferromagnetism can emerge in these heterostructures,
it is necessary to study the interplay of inversion-symmetry breaking, spin-orbit coupling, 
and correlation effects.

Motivated by these deliberations, we study in this article itinerant magnetic fluctuations in the Rashba-Hubbard model on the square lattice, 
which describes the salient features of interface electrons in a great number of 
heterostructures~\cite{rashba_kane_mele_rachel_PRB_14,zhang_NPJ_2015,brosco_capone_arXiv_18,ghadimi19,biderang19}
and which, moreover, is relevant for many noncentrosymmetric materials with 
strong spin-orbit coupling~\cite{riera_PRB_14,yanase_sigrist_JPSJ_07,yanase_sigrist_JPSJ_07b,yokoyama_tanaka_PRB_07}.
Previously, we have studied this model in the context of superconductivity using the random phase approximation (RPA), and found that both spin-singlet 
and spin-triplet superconductivity can arise~\cite{greco18}.
Here, we want to investigate the itinerant magnetism and study the magnetic fluctuations
as a function of electronic structure, on-site interaction $U$, and Rashba spin-orbit coupling (SOC). 
{In particular, we want to focus on the longitudinal ferromagnetic (FM) fluctuations, which occur for fillings
$n$ in between the two van Hove fillings, $n_{\textrm{vH$_2$}} < n < n_{\textrm{vH$_1$}}$. 
Our aim is to find the origin of these FM fluctuations and to show that they exist in a large region
of parameter space.

We find that the longitudinal FM fluctuations originate from interband contributions to the spin
susceptibility. These interband contributions are dominant if there is one holelike and
one electronlike Fermi surface (FS), i.e., when the filling $n$ is in between $n_{\textrm{vH$_2$}}$ and $n_{\textrm{vH$_1$}}$.
It follows from this insight, that longitudinal FM fluctuations occur quite commonly, i.e., 
in any Rashba system with one holelike and one electronlike Fermi surface. 
This is confirmed by our numerical calculations, which show  that FM fluctuations are
present whenever the filling is in between $n_{\textrm{vH$_2$}}$ and $n_{\textrm{vH$_1$}}$,
independent of the magnitude of the second-neighbor hopping and SOC.
The FM fluctuations  survive also up to values of $U$ close 
to the magnetic instability, as obtained within the RPA. 
We note that the mechanism for FM fluctuations presented in this article is markedly different
from Stoner ferromagnetism, which only occurs
close to large maxima in the density of states (DOS)~\cite{roemer_PRB_15}. 
As an experimental test to detect these type of FM fluctuations we propose to measure
the ratio between the longitudinal and transversal susceptibilities, which shows
pronounced features as a function of SOC and filling.}

The remainder of this paper is organized as follows. 
In Sec.~II we present briefly  our model and theoretical framework. In Sec.~III we 
study the itinerant fluctuations as a function 
of SOC and second-neighbor hopping $t'$. The origin of the ferromagnetic and antiferromagnetic fluctuations is discussed in Sec.~IV. In Sec.~V we propose possible
experimental tests for detecting the predicted ferromagnetic fluctuations. Section VI contains discussions and conclusions. In  
Appendices A and B we provide the main mathematical aspects of the present calculation. 

\section{Model and theoretical scheme}

The one-band Rashba-Hubbard model on the two-dimensional square lattice is defined by  
\begin{subequations} \label{def_rashba_hub}
\begin{equation} \label{rashba_hubbard}
H = \sum_{\bf k} {\psi}^\dag_{\bf k}  \hat{h} ( {\bf k} )    \psi^{\ }_{{\bf k}}  
+  U \sum_{{\bf k},{\bf k'}, {\bf q}} c^\dag_{{\bf k} \uparrow} c^{\ }_{{\bf k}+{\bf q} \uparrow}
c^\dag_{{\bf k'} \downarrow}c^{\ }_{{\bf k'}-{\bf q} \downarrow}, 
\end{equation}
where the single-particle Hamiltonian $\hat{h}({\bf k} )$ is 
\begin{equation}
\hat{h}({\bf k} ) = \left( \varepsilon_{\bf k} \tau_0 + {\bf g}_{\bf k} \cdot \boldsymbol{\tau}   \right).
\label{singleH}
\end{equation}
\end{subequations}
The band energy $\varepsilon_{\bf k}=-2t(\cos k_x +\cos k_y) + 4t' \cos k_x  \cos k_y -\mu$ 
contains both first- and second-neighbor hopping, $t$ and $t'$, respectively, and $\mu$ is 
the chemical potential~\cite{missprint}. The vector ${\bf g}_{\bf k}$ describes Rashba SOC with
${\bf g}_{\bf k}=V_{\textrm{so}}(\partial \varepsilon_{\bf k} / \partial k_y,  - \partial \varepsilon_{\bf k} / \partial k_x ,0)$
and the coupling constant $V_{\textrm{so}}$.
$\boldsymbol{\tau} = (\tau_1, \tau_2, \tau_3)^T$ are the three Pauli matrices
and $\tau_0$ stands for the $2\times2$ unit matrix. 
In Eq.~\eqref{rashba_hubbard}, $\psi_{\bf k}=(c_{{\bf k} \uparrow}, c_{{\bf k} \downarrow})^T$ is a doublet of annihilation
operators with wave vector ${\bf k}$ and  $U$ is the on-site Coulomb repulsion.  
In the following, energies are given in units of~$t$.

The presence of Rashba SOC splits
the electronic dispersion $\varepsilon_{\bf k}$ of the single-particle Hamiltonian~\eqref{singleH} into negative- and positive-helicity bands
with energies 
$E_{\bf k}^{1} = \varepsilon_{\bf k}  - \left| {\bf g}_{\bf k} \right|$ 
and $E_{\bf k}^{2} = \varepsilon_{\bf k}  + \left| {\bf g}_{\bf k} \right|$, 
respectively, see Fig.~\ref{fig1}(b).
These spin-split bands exhibit
a helical spin polarization, which is described
by the expectation value of the spin operator 
\begin{eqnarray} \label{eq_spin_pol}
\langle {\bf S}_{\bf k}  \rangle_i 
=
(-1)^i \frac{1}{2} \frac{ {\bf g}_{\bf k} } { |{\bf g}_{\bf k}|} ,
\end{eqnarray}
where $i$ denotes the band index. 
The spin polarization is proportional to the normalized ${\bf g}$-vector ${\bf g}_{\bf k}/|{\bf g}_{\bf k}|$, 
and thus is purely within the $xy$ plane. Moreover, the spin polarization is of helical nature, i.e., to a good approximation perpendicular to the momentum (tangential 
to the Fermi surface). 

In order to study the magnetic fluctuations of Hamiltonian~\eqref{def_rashba_hub}, we compute 
the spin susceptibility $\hat{\chi}  ({\bf q},i\omega_l)$ using RPA,
which is known to provide a reasonable description of the essential physics, at least within weak coupling~\cite{yanase_sigrist_JPSJ_07,yanase_sigrist_JPSJ_07b,yokoyama_tanaka_PRB_07,greco18}.
Within the RPA, the dressed spin susceptibility is given by 
\begin{equation} \label{eq_RPA}
\hat{\chi} ({\bf q},i\omega_l)
=
\left[ I -   \hat{\chi}^{(0)} ( {\bf q}, i \omega_l )   \hat{U} \right]^{-1} 
\hat{\chi}^{(0)} ({\bf q},i\omega_l) ,
\end{equation}
where $\hat{\chi}^{(0)}$ is the bare spin susceptibility.
Here, $\hat{\chi}$, $\hat{\chi}^{(0)}$, and $\hat{U}$ are $4 \times 4$ matrices containing the sixteen 
components of $\chi_{\sigma_1 \sigma_2 \sigma_3 \sigma_4}$,  
$\chi^{(0)}_{\sigma_1 \sigma_2 \sigma_3 \sigma_4}$, and $U$, respectively.
 The longitudinal and transversal susceptibilities can be
computed in terms of the matrix elements
$\chi_{\sigma_1 \sigma_2 \sigma_3 \sigma_4}$ as
\begin{subequations} \label{def_chi_trans_long}
\begin{equation} \label{def_chi_long}
\chi_{\textrm{long}}({\bf q},i\omega_l)=\chi_{\uparrow \uparrow \uparrow \uparrow}({\bf q},i\omega_l) - 
\chi_{\uparrow \downarrow \downarrow \uparrow} ({\bf q},i\omega_l),  
\end{equation}
\noindent and 
\begin{equation}
\chi_{\textrm{trans}}({\bf q},i\omega_l) = \chi_{\uparrow \uparrow \downarrow \downarrow}({\bf q},i\omega_l), 
\end{equation}
\end{subequations}
\noindent respectively.
More details on the derivation of the dressed spin susceptibility~\eqref{eq_RPA} are given in Appendix~\ref{appendix_RPA}.

\section{Spin fluctuations of Rashba-Hubbard model}
\label{results_section}

\begin{figure}[t!]
\centering
\includegraphics*[width=8cm,angle=0]{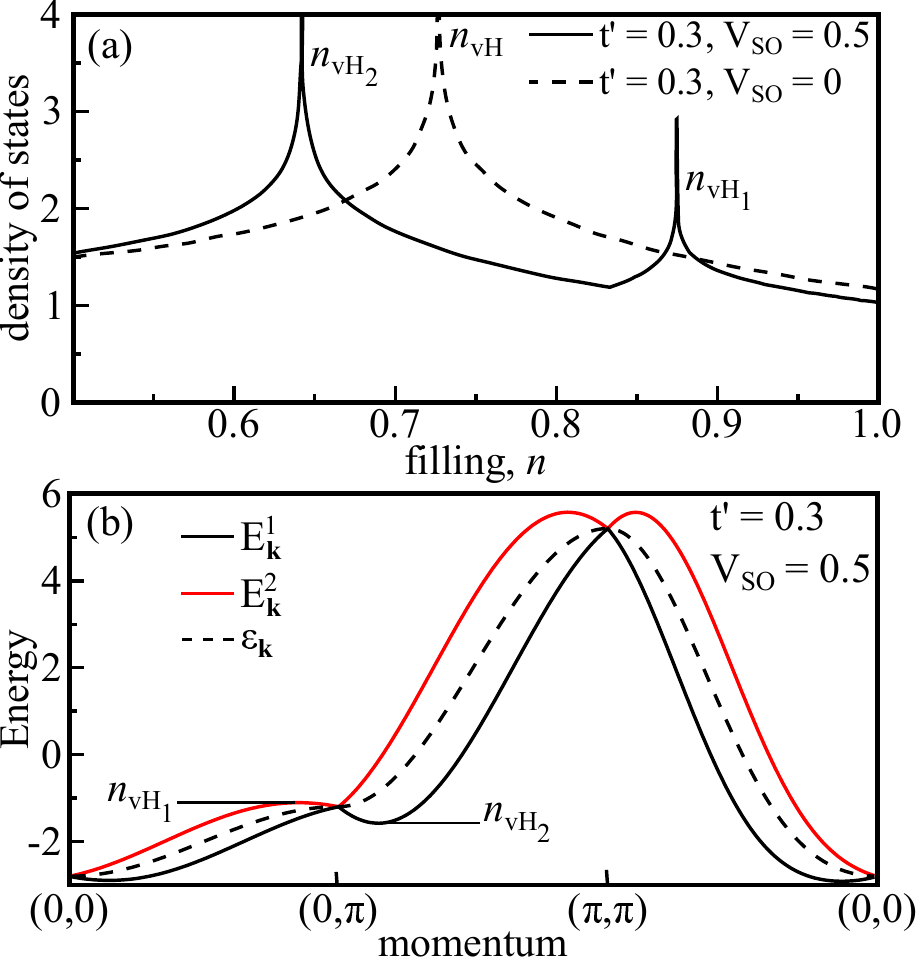}
\caption{
(a) Density of states versus filling $n$ for $t'=0.3$, with $V_{\textrm{so}}=0$ (dashed line) and 
$V_{\textrm{so}}=0.5$ (solid line). 
(b)~Band dispersions for $t'=0.3$, with $V_{\textrm{so}}=0$ (dashed line) and 
$V_{\textrm{so}}=0.5$ (solid lines). 
}
\label{fig1}
\end{figure}

To set the stage, we first recall some properties of the spin fluctuations in the square-lattice Hubbard model without SOC but finite $t'$,
corresponding to $V_{\textrm{so}}=0$ in Eq.~\eqref{def_rashba_hub}. 
In the absence of SOC, full $SU(2)$ spin-rotation symmetry is preserved in the paramagnetic phase,
and hence the longitudinal and transversal spin susceptibilities are equal, i.e., $\chi_{\textrm{long}}=\chi_{\textrm{trans}}$.
As has been shown in numerous works~\cite{kamogawa19,roemer_PRB_15,roemer_arXiv_2019,fleck97},
the spin fluctuations are in this case mostly of (incommensurate) AFM nature. 
Only very close to the van-Hove filling $n_{\textrm{vH}}$ there occur
ferromagnetic fluctuations, which diminish quickly for fillings away from $n_{\textrm{vH}}$.
These FM fluctuations can be understood as resulting
from Stoner ferromagnetism~\cite{roemer_PRB_15,kamogawa19}, which occurs for fillings close
to a large asymmetric maximum in the DOS. Indeed,
for finite $t'$, the maximum in the DOS at the van-Hove filling $n_{\textrm{vH}}$
is always asymmetric  [see dashed line in Fig.~\ref{fig1}(a)], such that the
Stoner criterion for ferromagnetism can be satisfied. 
We note, however, that for vanishing $t'$ the DOS is symmetric, such that the Stoner criterion cannot
be fulfilled. Hence, for $t'=0$ the fluctuations are AFM also close to the van-Hove filling, due to perfect nesting of the Fermi surface.

\begin{figure}[t!]
\centering
\includegraphics[width=8cm,angle=-0]{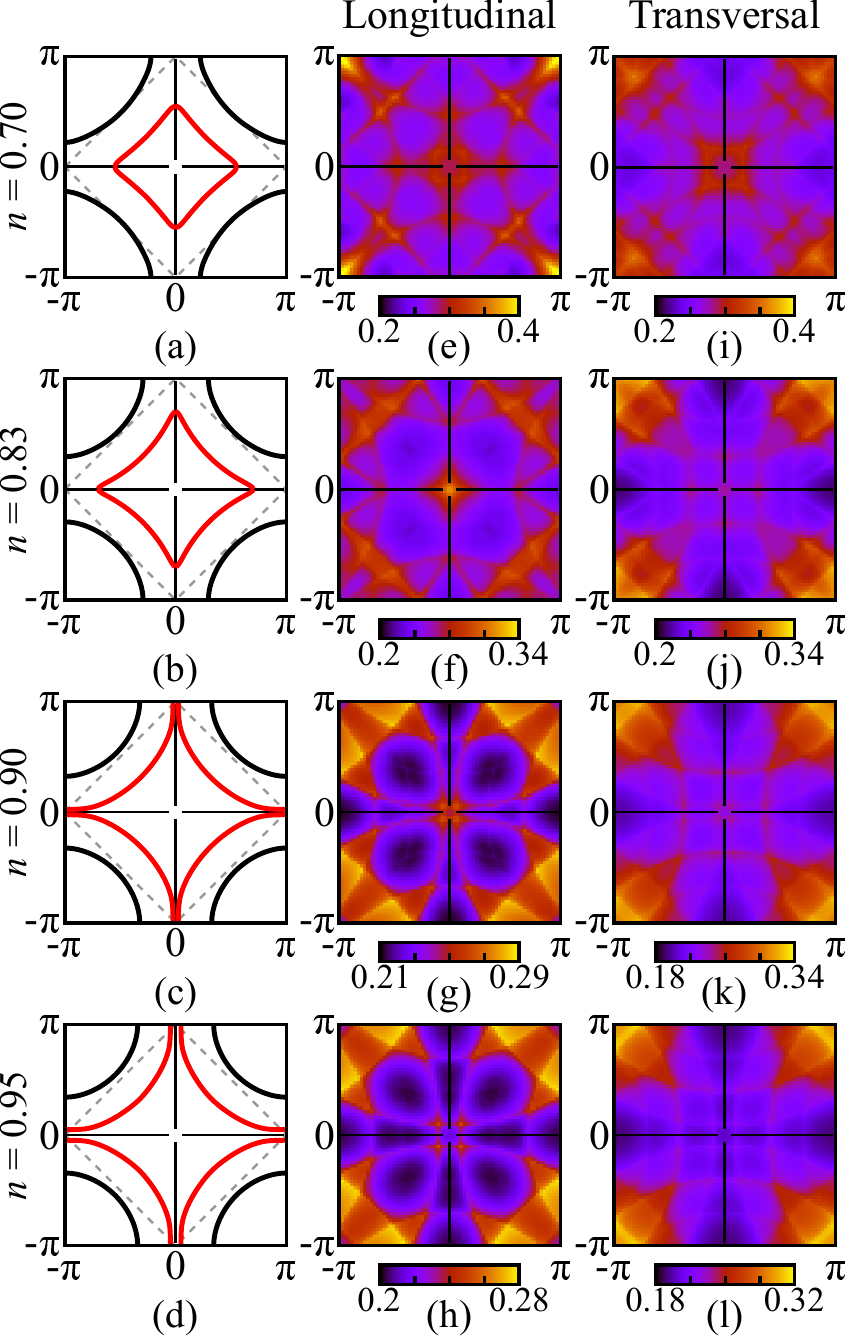}
\caption{ \label{mapLT} \label{fig2}
(a)-(d) Fermi surface topology for four different fillings.
{The dashed lines indicate the AFM zone boundary.}
(e)-(l)
Color maps of the bare static susceptibilities
as a function of modulation vector ${\bf q}$ for each of the four fillings.
The second and third columns show the longitudinal and transversal static susceptibilities,
${\chi}^{(0)}_{\textrm{long}} ({\bf q} )$ and ${\chi}^{(0)}_{\textrm{trans}} ({\bf q} )$,
respectively. 
Here, we set the temperature to 
$T=0.01$, and choose $t'=0.3$ and $V_{\textrm{so}}=0.5$, for which the two van Hove singularities
are located at $n_{\textrm{vH$_1$}}=0.87$ and $n_{\textrm{vH$_2$}}=0.65$.
}
\end{figure}

For finite Rashba SOC ($V_{\textrm{so}} \ne 0$) the situation changes drastically. 
First of all, Rashba SOC lifts the spin degeneracy of the bands,
thereby splitting the van-Hove singularity into two divergences  that
occur at the fillings $n_{\textrm{vH$_1$}}$ and $n_{\textrm{vH$_2$}}$, see solid lines in Fig.~\ref{fig1}.
These two van-Hove singularities originate from saddle points
in the dispersion at $(0, \pi - \delta)$, $(\delta, \pi)$, and symmetry related points.
At these saddle points the gradient of the dispersion vanishes, causing   logarithmic divergences
in the DOS.
Importantly, the topology of the Fermi surfaces changes as the filling $n$ crosses
the two van-Hove fillings: For $n > n_{\textrm{vH$_1$}}$ the two Fermi surfaces are hole-like and centered arount $(\pi, \pi)$, see Figs.~\ref{fig2}(c) and~\ref{fig2}(d).
For $n_{\textrm{vH$_2$}} < n < n_{\textrm{vH$_1$}}$, on the other hand, one Fermi surface is electron-like, while the other his hole-like [Figs.~\ref{fig2}(a) and~\ref{fig2}(b)].
For $n < n_{\textrm{vH$_2$}}$, both Fermi surfaces are electron-like and centered around $(0,0)$.

As is known from a great many works on the Hubbard model~\cite{roemer_PRB_15,scalapino_hirsch_86,hlubina_PRB_99}, the Fermi surface topology strongly influences
the structure of the spin fluctuations. To uncover this dependence, we plot
in Figs.~\ref{fig2}(e)-\ref{fig2}(l) the bare static susceptibility $\hat{\chi}^{(0)} ({\bf q}, \omega = 0) \equiv \hat{\chi}^{(0)} ({\bf q} )$ 
as a function of modulation vector ${\bf q}$ for different fillings $n$. 
We find that for fillings with two hole-like Fermi surfaces ($n > n_{\textrm{vH$_1$}}$), 
the dominant modulation vector of the longitudinal spin susceptibility  ${\chi}^{(0)}_{\textrm{long}} ({\bf q} )$ is incommensurate AFM.
For fillings with one electron-like and one hole-like Fermi surface ($n_{\textrm{vH$_2$}} < n < n_{\textrm{vH$_1$}}$), however,
the longitudinal spin fluctuations are mostly FM [see Fig.~\ref{fig2}(f)]. 
Finally, for fillings with two electron-like Fermi surfaces ($n < n_{\textrm{vH$_2$}}$), the
longitudinal spin fluctuations are dominantly AFM (not shown).
The transversal spin susceptibility  ${\chi}^{(0)}_{\textrm{trans}} ({\bf q} )$, in contrast to the longitudinal one,
shows (incommensurate) AFM fluctuations for almost   all fillings $n$.

\begin{figure}[t!]
\centering
\includegraphics*[width=8cm,angle=0]{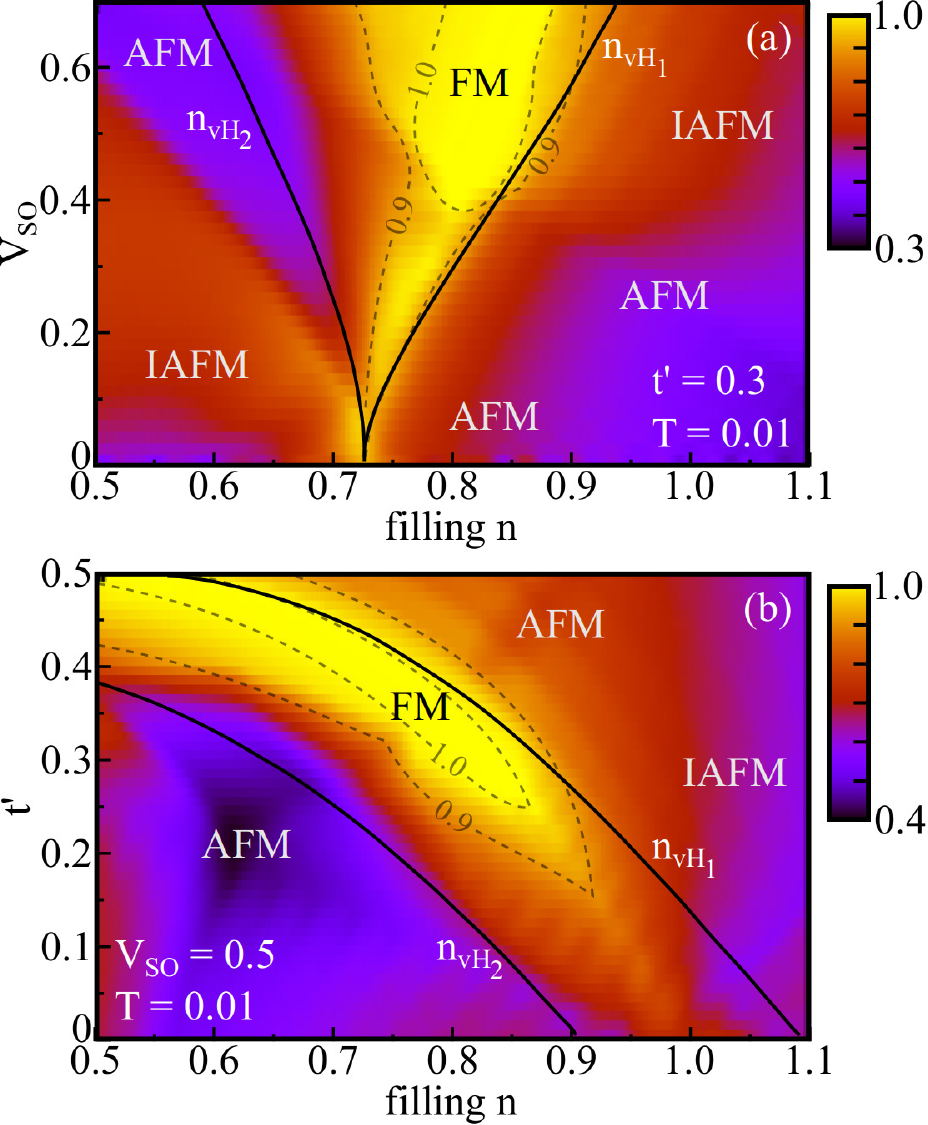}
\caption{  \label{fig3}
Relative intensity of the FM fluctuations compared to the incommensurate AFM fluctuations in the longitudinal susceptibility,
${\chi}^{(0)}_{\textrm{long}} ({\bf q} = {\bf 0}  ) /  {\chi}^{(0)}_{\textrm{long}} ( \tilde{\bf q} ) $,
as a function of filling $n$, Rashba SOC $V_{\textrm{so}}$ [panel  (a)], and second neighbor hopping $t'$ [panel  (b)].
The solid lines represent the two van Hove fillings $n_{\textrm{vH$_1$}}$ and $n_{\textrm{vH$_2$}}$.
IAFM stands for incommensurate AFM.
}
\label{FMW}
\end{figure}

These findings are independent of the particular values of $t'$ and $V_{\textrm{so}}$,
as shown in Fig.~\ref{fig3}. {Here, we present the relative intensities 
of the FM fluctuations compared to the (incommensurate) AFM fluctuations in the
longitudinal susceptibility. I.e., we plot
${\chi}^{(0)}_{\textrm{long}} ({\bf q} = {\bf 0}  ) /  {\chi}^{(0)}_{\textrm{long}} ( \tilde{\bf q} ) $
where $ \tilde{\bf q} $ is the location of the maximum of ${\chi}^{(0)}_{\textrm{long}}  $. 
Inside the contour labeled ``1" the maximum of ${\chi}^{(0)}_{\textrm{long}}  $ is
at the FM vector $\tilde{\bf q}=0$, while outside this contour, the maximum is at
some (incommensurate) AFM vector, i.e., (close to) at $ \tilde{\bf q} = (\pi, \pi)$. 
In Fig.~\ref{fig3}(a) we plot the ratio ${\chi}^{(0)}_{\textrm{long}} ({\bf q} = {\bf 0}  ) /  {\chi}^{(0)}_{\textrm{long}} ( \tilde{\bf q} ) $ }
as a function of  $V_{\textrm{so}}$ and filling $n$ with fixed $t'=0.3$, while in Fig.~\ref{fig3}(b)  it is shown
as a function of $t'$ and filling $n$ with fixed $V_{\textrm{so}}=0.5$. 
We observe a broad region, marked in yellow, where dominant FM fluctuations occur. These
regions are bounded by the two van Hove fillings $n_{\textrm{vH$_1$}}$ and $n_{\textrm{vH$_2$}}$ (black lines).
{The full width at half maximum of the FM peak in these regions is about $0.4$, corresponding
to a correlation length of about fifteen lattice constants.}
In Fig.~\ref{fig3}(a) we see that with decreasing   $V_{\textrm{so}}$, the FM fluctuation region becomes narrower and narrower, 
and shrinks to a single point at  $n_{\textrm{vH}} = n_{\textrm{vH$_1$}}=n_{\textrm{vH$_2$}}$ for $V_{\textrm{so}}=0$.
From Fig.~\ref{fig3}(b) we find that as $t'$ is increased the FM fluctuations occur at lower fillings $n$.
Moreover, the  FM fluctuations become dominant only for $t'$ larger than a certain onset value, i.e., for $t'\gtrsim 0.2$.

We note that in the entire parameter space the spin fluctuations are either FM (peaked at ${\bf q}=0$),
commensurate AFM (peaked at ${\bf q}=(\pi, \pi)$), or incommensurate AFM (peaked at $\tilde{\bf q}$, with 
$\tilde{\bf q}$ away from, but close to $(\pi, \pi)$). Hence, the transition from FM to (incommensurate) AFM does not occur
smoothly via a continuous evolution of the modulation vector ${\bf q}$, but rather abruptly when
the peak at $\tilde{\bf q}$ suddenly becomes larger than the one at ${\bf q}=0$.

\begin{figure}[t!]
\centering
\includegraphics*[width=8.6cm]{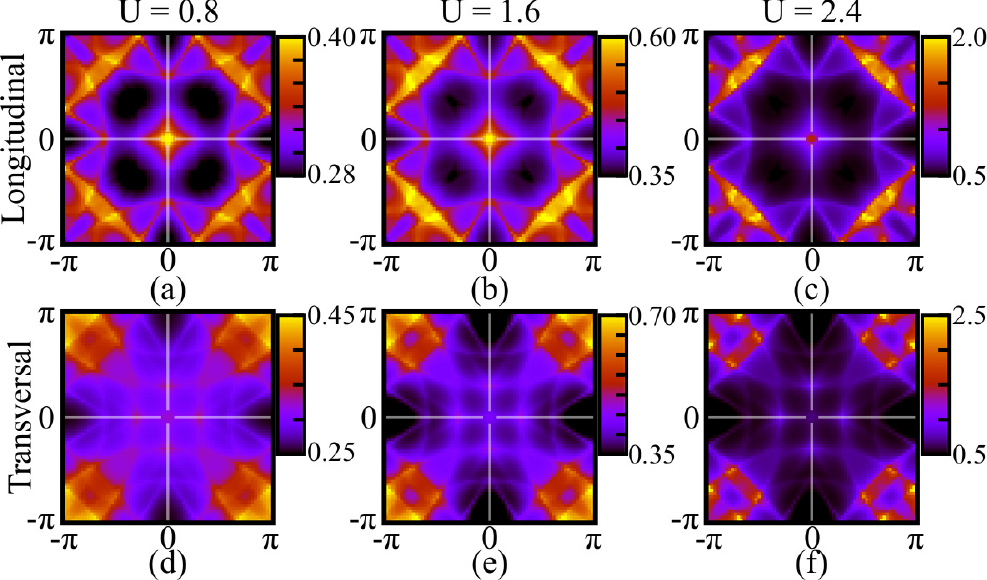}
\caption{ \label{fig_dressed_spin_suscep}
Color maps of the dressed static susceptibilities as a function of ${\bf q}$ for filling $n=0.83$ and different onsite interactions $U$ approaching the critical interaction $U_c$.  
The first and second rows show the longitudinal and transversal static susceptibilities, $\chi_{\textrm{long}} ( {\bf q} )$
and $\chi_{\textrm{trans}} ( {\bf q} )$, respectively. 
As in Fig.~\ref{fig2}, we set $T=0.01$, $t'=0.3$, and $V_{\textrm{so}} =0.5$, in which case
  $U_c \simeq 2.5$.
}
\end{figure}

So far, we have focused on the bare susceptibility $\hat{\chi}^{(0)} ({\bf q} )$.  The spin fluctuations  of the dressed 
spin susceptibility $\hat{\chi}  ({\bf q} )$, Eq.~\eqref{eq_RPA},
are shown in Fig.~\ref{fig_dressed_spin_suscep}. For small and intermediate $U$ the structure of the spin fluctuations
of $\hat{\chi}  ({\bf q} )$ is {almost identical} to the spin fluctuations of  $\hat{\chi}^{(0)} ({\bf q} )$.
{This is to be expected, since a purely onsite interaction cannot change the momentum dependence of the spin fluctuations.}
For low and high fillings, $n < n_{\textrm{vH$_2$}}$ and $n> n_{\textrm{vH$_1$}}$, both   longitudinal and transversal
susceptibilities show dominant incommensurate AFM fluctuations.
In between the two van Hove fillings, $n_{\textrm{vH$_2$}}  < n < n_{\textrm{vH$_1$}}$, the transversal  susceptibility
exhibits incommensurate AFM fluctuations, while the longitudinal one shows FM fluctuations. 
{These findings do not depend on the particular values of $t'$ and $V_{\textrm{so}}$.
That is, the phase diagram of Fig.~\ref{FMW}, which shows the boundaries between the different magnetic fluctuations,
remains almost identical upon inclusion of a small or intermediate onsite interaction $U$.
Increasing $U$ beyond intermediate values, the FM fluctuations 
rapidly decrease as the {critical interaction} $U_{\textrm{c}} \simeq 2.5$ is approached~\cite{greco18},  see first row of Fig.~\ref{fig_dressed_spin_suscep}.
For strong interactions $U \sim U_c$ (incommensurate) AFM fluctuations dominate 
[see Fig.~\ref{fig_dressed_spin_suscep}(c) and (f)], leading to AFM order with magnetic moments oriented in-plane.

Hence, we conclude that the magnetic fluctuations remain largely unchanged by onsite interactions $U$ of small and intermediate strength. 
In particular, the FM fluctuations are unaffected; they originate from finite $t'$ and finite SOC, rather than the interaction $U$.
Thus, in order to uncover the root of the FM and AFM fluctuations, it is sufficient to consider the bare susceptibility
$\hat{\chi}^{(0)} ({\bf q} )$, whose form is known exactly, and which can be analyzed using analytical means.
This is the purpose of the next section.}

\section{Origin of magnetic fluctuations}
\label{sec_origin_FM}

{In this section we want to study the origin of the longitudinal FM and AFM fluctuations.
To do so, we can focus on the bare susceptibility $\hat{\chi}^{(0)}_{\textrm{long}}$, as discussed above.}
We observe that $\hat{\chi}^{(0)}_{\textrm{long}}$ can
be separated into interband and intraband parts. That is, 
\begin{eqnarray} \label{sucep_intra_inter}
\chi^{(0)}_{\textrm{long}} ( {\bf q}, i \omega_l )
=
 \chi^{(0), \textrm{intra}}_{\textrm{long}}( {\bf q}, i \omega_l ) + \chi^{(0), \textrm{inter}}_{\textrm{long}}( {\bf q}, i \omega_l ),
\end{eqnarray}
\noindent where $\chi^{(0), \textrm{intra}}_{\textrm{long}}$ and $\chi^{(0), \textrm{inter}}_{\textrm{long}}$
are given in Appendix~\ref{appendix_B}. 
As discussed in Appendix~\ref{appendix_B}, the AFM fluctuations originate from the intraband term,
while the FM fluctuations stem from the interband term, see Eqs.~\eqref{myEq_B3} to~\eqref{myEq_B6}.

\begin{figure}[t!]
\centering
\includegraphics*[width=7cm,angle=-0]{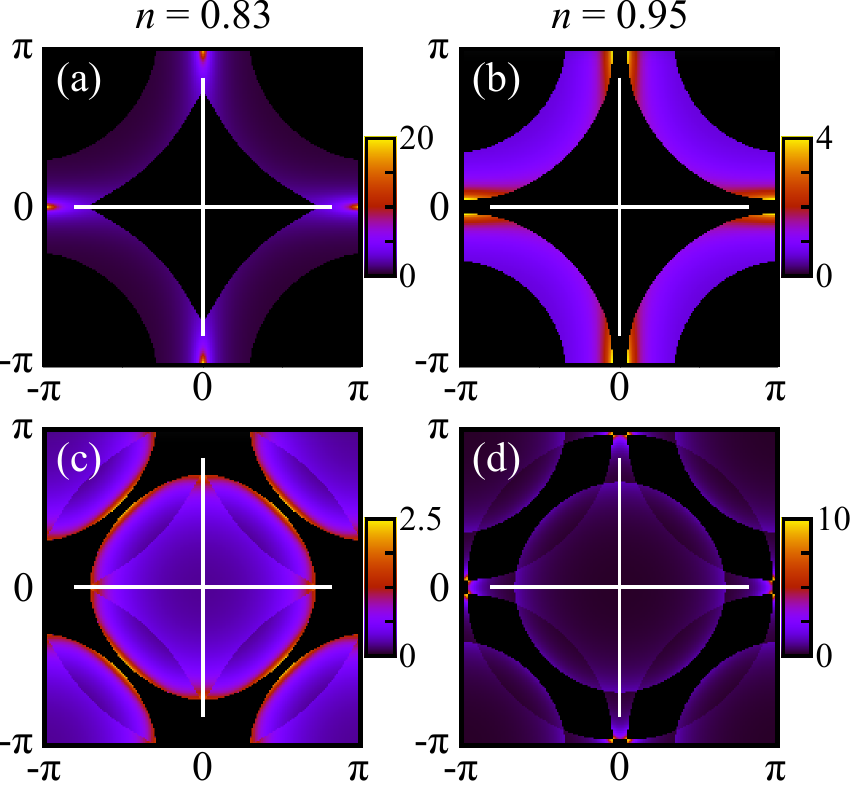}
\caption{
{Momentum dependence of the summand of Eq.~\eqref{chi_long_q_0}
[(a) and (b)] and of the summand of  Eq.~\eqref{chi_long_q_pi_pi} [(c) and (d)].
In (a), (c) the filling is $n=0.83$, while in (b), (d) it is $n=0.95$.
The parameters are $t'=0.3$, $V_{\textrm{so}}=0.5$, and $\Gamma = 0.05$.
Note that four different color scales are used.}
}
\label{summand}
\end{figure}

{Let us first discuss the interband term, which is responsible for FM fluctuations.
At the FM modulation vector ${\bf q} = (0,0)\equiv {\bf 0}$, the static interband susceptibility $\chi^{(0), \textrm{inter} }_{\textrm{long}}  ( {\bf 0},  \omega = 0 ) $ takes the simple form
\begin{eqnarray} \label{chi_long_q_0}
\chi^{(0), \textrm{inter} }_{\textrm{long}}  ( {\bf 0}  ) 
=
 \sum_{ {\bf k}} \left[f(E_{\bf k}^{1})-f(E_{\bf k}^{2}) \right]    \frac{4|{g}_{\bf k}|}{4|{g}_{\bf k}|^2+ \Gamma^2}  , \; \; 
\end{eqnarray} 
where $\Gamma $ is a small positive infinitesimal.
Because $E_{\bf k}^{1} < E_{\bf k}^{2}$ and $f(z)$ is a decreasing function of $z$, we have 
$0 \leq f(E^1_{\bf k} ) - f ( E^2_{\bf k} ) \leq 1$ in the above expression. 
In the limit $\Gamma \to 0$, the summand in Eq.~\eqref{chi_long_q_0} exhibits a divergence at those ${\bf k}$ where
${\bf g}_{\bf k} = 0$ and $f(E^1_{\bf k} ) - f ( E^2_{\bf k} )$ is nonzero. 
 Since ${\bf g}_{\bf k}$ is inversion antisymmetric in ${\bf k}$, it vanishes at the four inversion-invariant momenta
\begin{eqnarray} \label{inversion_inv_momenta}
{\bf k} \in \{ (0,0), (0, \pi), (\pi,0), (\pi, \pi ) \} .
\end{eqnarray}
Hence, the second factor of Eq.~\eqref{chi_long_q_0} becomes larger and larger, and eventually diverges,
as the above four momenta are approached.
The Fermi factor $ f(E_{\bf k}^{1})-f(E_{\bf k}^{2}) $, on the other hand,
also always vanishes at the four momenta of Eq.~\eqref{inversion_inv_momenta},
where $E^1_{\bf k} = E^2_{\bf k}$. However, it can be nonzero in an arbitrarily small neighborhood
around these points. This occurs near ${\bf k} \in \{(0, \pi), (\pi,0) \}$, when one Fermi surface is electron-like and the other one is hole-like.
If the two Fermi surfaces are both electron-like (or hole-like), then the Fermi factor vanishes in a finite neighborhood around
all four momenta of Eq.~\eqref{inversion_inv_momenta},
thereby cancelling the divergence of the second factor in Eq.~\eqref{chi_long_q_0}.
These findings are illustrated in Figs.~\ref{summand}(a) and~\ref{summand}(b), which show the momentum dependence
of the summand of Eq.~\eqref{chi_long_q_0}. For filling $n=0.83$, corresponding to one electron-like 
and one hole-like Fermi surface, we observe that the summand diverges near ${\bf k} \in \{(0, \pi), (\pi,0) \}$, 
while for filling $n=0.95$, corresponding to two hole-like Fermi surfaces, the summand
is finite for all ${\bf k}$.
From this we deduce that strong FM fluctuations  
occur  only for fillings in between the two van  Hove fillings, i.e., when
one Fermi surface is electron-like and the other one hole-like. 
As an aside, we note that 
close to the second van Hove singularity $n_{vH_2}$, i.e., for fillings $n_{vH_2} + \epsilon$ (with $\epsilon >0$) there is an additional Fermi pocket
around $(0,\pi)$ and $(\pi, 0)$.
This additional Fermi pocket renders the Fermi factor $f(E^1_{\bf k})- f(E^2_{\bf k} )$ zero in the neighborhood of $(0, \pi)$ and $(\pi, 0)$.
Therefore, the FM fluctuations, that originate from divergences of the second factor of Eq.~\eqref{chi_long_q_0} at $(0, \pi)$ and $(\pi, 0)$, are suppressed
for fillings close to $n_{vH_2}$, see Fig.~\ref{fig3}.
}

{Next we study the intraband term of Eq.~\eqref{sucep_intra_inter}, which produces AFM fluctuations.
At the AFM modulation vector ${\bf q} = ( \pi, \pi ) \equiv  {\bf Q}$,
the static intraband susceptibility $\chi^{(0), \textrm{intra} }_{\textrm{long}}  ( {\bf Q},  \omega = 0 ) $ takes the simple form
\begin{eqnarray} \label{chi_long_q_pi_pi}
&&  \chi^{(0), \textrm{intra}}_{\textrm{long}}( {\bf Q} )
=
\\
&& \quad
\sum_{\alpha=1,2}
\sum_{ {\bf k}} 
\left[
f(E_{\bf k+Q}^{\alpha})-f(E_{\bf k}^{\alpha})
\right]
\frac{
E_{\bf k+Q}^{\alpha}-E_{\bf k}^{\alpha}
}
{
(E_{\bf k+Q}^{\alpha}-E_{\bf k}^{\alpha} )^2 + \Gamma^2
} .
\nonumber
\end{eqnarray}
In the limit $\Gamma \to 0$, the summand in Eq.~\eqref{chi_long_q_pi_pi} has a divergence at those ${\bf k}$ where
$E_{\bf k+Q}^{\alpha} = E_{\bf k}^{\alpha} $ and $f(E_{\bf k+Q}^{\alpha})-f(E_{\bf k}^{\alpha})$ is nonzero. 
The condition $E_{\bf k+Q}^{\alpha} = E_{\bf k}^{\alpha} $ is satisfied at the AFM zone boundary, indicated
by the dashed lines in Figs.~\ref{fig2}(a) - \ref{fig2}(d). 
This leads to a divergence of the second factor of Eq.~\eqref{chi_long_q_pi_pi},
which is further enhanced near $(0, \pi)$ and $(\pi,0)$ by the saddle points
in the dispersions of $E_{\bf k+Q}^{\alpha}$ and $E_{\bf k}^{\alpha}$.
The Fermi factor $f(E_{\bf k+Q}^{\alpha})-f(E_{\bf k}^{\alpha})$, on the other hand,
is nonzero near  $(0, \pi)$ and  $(\pi,0)$, only
for the second band with fillings $n >  n_{\textrm{vH$_1$}}$ and
for the first band with fillings $n < n_{\textrm{vH$_2$}}$.
For $ n_{\textrm{vH$_2$}}<   n < n_{\textrm{vH$_1$}}$, however,
the Fermi factor always vanishes near $(0, \pi)$ and  $(\pi,0)$,
thus cancelling the divergence from the second factor in Eq.~\eqref{chi_long_q_pi_pi}. 
These observations are illustrated by Figs.~\ref{summand}(c) and~\ref{summand}(d), which 
display the ${\bf k}$ dependence of the summand of Eq.~\eqref{chi_long_q_pi_pi}. 
 For $n=0.95$, corresponding to two hole-like Fermi surfaces,
 the summand show divergences near $(0, \pi)$ and  $(\pi,0)$,
 while for $n=0.83$, corresponding to one hole-like and one electron-like Fermi surface,
 the summand does not show any  divergences.
We conclude that strong AFM fluctuations occur only 
for $n < n_{\textrm{vH$_2$}}$ and $n >  n_{\textrm{vH$_1$}}$,
but not in between the two van Hove fillings.

In this section we have focused on the bare susceptibility $\hat{\chi}^{(0)}_{\textrm{long}}$. But the above arguments also explain the
origin of the FM and AFM fluctuations in the dressed susceptibility $\hat{\chi}^{\ }_{\textrm{long}}$, since an onsite interaction $U$ of
small or intermediate strength does not alter the  structure of the magnetic fluctuations (see discussion at the end of Sec.~\ref{results_section}).
It is possible to generalize the given arguments  in a straightforward manner  to other Rashba systems on 
other types of lattices with correlations of weak or intermediate strength.
Thus, we expect that FM fluctuations occur generically
for a large class of Rashba systems with one electron-like and one hole-like Fermi surface.  }

 \section{Experimental test to identify ferromagnetic fluctuations}

In order to identify the discussed FM fluctuations in experiments, we propose to
measure the ratio between the longitudinal and transversal static susceptibilities
 in the presence 
of a constant magnetic field, i.e., 
$R   = \chi_{\textrm{long}} ({\bf 0})  / \chi_{\textrm{trans}} ({\bf 0}) $.
In an experiment $\chi_{\textrm{long}} ({\bf 0})$ is the response 
to a constant magnetic field perpendicular to the two-dimensional layer, 
while $ \chi_{\textrm{trans}} ({\bf 0})$ is the response to a field parallel to the layer. 
The ratio  $R$  
is expected to depend  only weakly on material details.
Moreover, $R$ shows pronounced features as a function of filling $n$ and Rashba SOC $V_{\textrm{so}}$,
for which one could look for in experiments.

\begin{figure}[t!]
\centering
\includegraphics*[width=9.5cm,angle=-0]{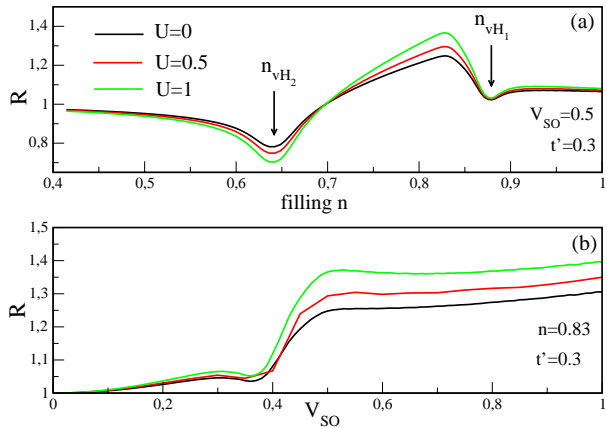}
\caption{ 
Ratio between the longitudinal and transversal static susceptibilities
$R  = \chi_{\textrm{long}} ({\bf 0})  / \chi_{\textrm{trans}} ({\bf 0})$ as a function
of (a) filling $n$ and (b) Rashba SOC $V_{\textrm{so}}$ for different values of~$U$.
In (a) the parameters are $t'=0.3$ and $V_{\textrm{so}}=0.5$, 
while in (b) we set $t'=0.3$ and $n=0.83$.
}
\label{ratio}
\end{figure}

In Fig.~\ref{ratio}(a) we present the results for $R$ versus filling~$n$ for $V_{\textrm{so}}=0.5$, $t'=0.3$, and different values of $U$.    
The broad peak larger than $1$  for fillings in between the two van Hove fillings, $n_{\textrm{vH$_2$}} < n < n_{\textrm{vH$_1$}}$,
originates from the dominant FM fluctuations in $\chi_{\textrm{long}}$, as discussed above. 
{As a function of onsite interaction $U$, the height of this peak remains nearly unchanged,
for small and intermediate values of $U$. 
For strong $U$ close to $U_c \simeq 2.5$, 
the FM fluctuations rapidly vanish, as discussion in Sec.~\ref{results_section}. }

In Fig.~\ref{ratio}(b) we plot $R$  as a function of Rashba SOC $V_{\textrm{so}}$ for the filling $n=0.83$ , at which
the FM fluctuations are strongest. The values of $U$ are the same as in panel~(a).
Interestingly, $R$ versus  $V_{\textrm{so}}$ shows a pronounced step at  $V_{\textrm{so}}^{\textrm{onset}}$, 
above which longitudinal FM fluctuations occur, cf.~Fig.~\ref{fig3}(a).  
We note that  $V_{\textrm{so}}^{\textrm{onset}}$ corresponds to the SOC strength, for which
the second van Hove singularity is located at the filling  $n=0.83$, i.e., entering the yellow region in Fig.~\ref{fig3}(a).
As in Fig.~\ref{ratio}(a), we find that $R$ above the step does not change much with increasing $U$, as long
as $U$ is smaller than   $U_c $.

The discussed dependence of $R$ on filling $n$ and Rashba SOC $V_{\textrm{so}}$ could   be measured in heterostructure interfaces.
In these interfaces it is possible to control the filling by doping or gating~\cite{thiel_mannhart_LAO_STOscience_06}. The Rashba SOC, 
on the other hand, can be tuned by applying electric fields or by modulating the layer thickness~\cite{caviglia_triscone_Nature_08,matsuda_PRL_14}.  

\section{Discussions and conclusions}

In summary, we have studied magnetic fluctuations of the $t$-$t'$-Rashba-Hubbard model on the square lattice.
The Rashba spin-orbit coupling of this model splits the bands and leads to two 
van Hove singularities. 
We have found that for fillings in between these two van Hove singularities, there exist 
dominant ferromagnetic fluctuations in the longitudinal susceptibility.
Outside this filling region  the magnetic fluctuations are (incommensurate) antiferromagnetic. 
{The ferromagnetic fluctuations remain largely unchanged
by onsite interactions $U$ of small and intermediate strength.}
They originate from
interband contributions to the longitudinal susceptibility.
These interband contributions only exist if there is a hole-like and an electron-like Fermi surface,
which is the case for fillings in between the two van Hove singularities. 
Thus, the origin of these ferromagnetic fluctuations  is 
 different from the Stoner 
criterion for ferromagnetism, which is only satisfied close to large maxima in the density of states. 
As discussed in Sec.~\ref{sec_origin_FM}, 
these type of ferromagnetic fluctuations 
are expected to occur more generally, i.e., in
any Rashba system with one electron-like and one hole-like Fermi surface.  

We hope that our findings will stimulate experimentalists to look for two-dimensional materials 
or noncentrosymmetric systems that satisfy 
these conditions.
In order to identify the ferromagnetic fluctuations in an experiment, 
we have proposed to measure the ratio between the longitudinal 
and transversal susceptibilities. This ratio 
is expected to depend only weakly on material details.
It shows a pronounced step as a function
of spin-orbit coupling and a broad peak as a function of filling, see Fig.~\ref{ratio}. 

To conclude, we mention several directions for future research. 
First of all, the reported ferromagnetic fluctuations could provide 
a pairing mechanism for unconventional superconductivity.
We have recently reported some initial results concerning this in Ref.~\cite{greco18}.
It would  be interesting to study in more detail the pairing symmetry
of the superconductivity that is induced by the ferromagnetic fluctuations.
Furthermore, it would be worthwhile to investigate the magnetic fluctuations
of the Rashba-Hubbard model using more advanced techniques, such as FLEX~\cite{YANASE20031}
or fRG~\cite{fRG_review_metzner_RMP_12}. Within the RPA we find that the ferromagnetic fluctuations do not lead
to ferromagnetic order, since the antiferromagnetic fluctuations become stronger
as $U$ approaches $U_{\textrm{c}}$. It would be interesting to know, whether
this result is confirmed by more sophisticated methods.

\acknowledgments
We thank G.~Jackeli, G.~Logvenov, H.~Nakamura, A.M.~Ole\'{s}, and M.~Sigrist for useful discussions. 
A.G.~thanks Max Planck Institute for Solid State Research in Stuttgart for 
hospitality and financial support.

\appendix
\section{Spin susceptibility within the \\ random phase approximation} 
\label{appendix_RPA}

In this appendix, we give the precise definition of the  dressed spin susceptibility.
Within the RPA the dressed spin susceptibility $\chi_{\sigma_1 \sigma_2 \sigma_3 \sigma_4} ({\bf q},i\omega_l)$ 
is given by~\cite{yanase_sigrist_JPSJ_07,yanase_sigrist_JPSJ_07b,greco18}
\begin{equation} \label{eq_RPA_suscep}
\hat{\chi} ({\bf q},i\omega_l)
=
\left[ I -   \hat{\chi}^{(0)} ( {\bf q}, i \omega_l )   \hat{U} \right]^{-1} 
\hat{\chi}^{(0)} ({\bf q},i\omega_l), 
\end{equation}
where $\hat{\chi}^{(0)}$ is the bare spin susceptibility and $\hat{U}$ is the interaction matrix. Here, $\hat{\chi}^{(0)}$ and $\hat{\chi}$ are $4 \times 4$ matrices
with the matrix elements $\chi^{(0)}_{\sigma_1 \sigma_2 \sigma_3 \sigma_4}$ and $\chi_{\sigma_1 \sigma_2 \sigma_3 \sigma_4}$, respectively. 
The explicit form of the matrix $\hat{\chi}^{(0)}$ is given by
\begin{equation}
\hat{\chi}^{(0)}= 
\left(
\begin{array}{llll}
\chi^{(0)}_{\uparrow \uparrow \uparrow \uparrow}& \chi^{(0)}_{\uparrow \downarrow \uparrow \uparrow}& \chi^{(0)}_{\uparrow \uparrow \downarrow \uparrow}
&\chi^{(0)}_{\uparrow \downarrow \downarrow \uparrow}\\
\chi^{(0)}_{\uparrow \uparrow \uparrow \downarrow}& \chi^{(0)}_{\uparrow \downarrow \uparrow \downarrow}& \chi^{(0)}_{\uparrow \uparrow \downarrow \downarrow}
&\chi^{(0)}_{\uparrow \downarrow \downarrow \downarrow}\\
\chi^{(0)}_{\downarrow \uparrow \uparrow \uparrow}& \chi^{(0)}_{\downarrow \downarrow \uparrow \uparrow}& \chi^{(0)}_{\downarrow \uparrow \downarrow \uparrow}
&\chi^{(0)}_{\downarrow \downarrow \downarrow \uparrow}\\
\chi^{(0)}_{\downarrow \uparrow \uparrow \downarrow}& \chi^{(0)}_{\downarrow \downarrow \uparrow \downarrow}& \chi^{(0)}_{\downarrow \uparrow \downarrow \downarrow}
&\chi^{(0)}_{\downarrow \downarrow \downarrow \downarrow}
\end{array}
\right) \; ,
\end{equation}
and similarly for $\hat{\chi}$.
The interaction matrix $\hat{U}$  is a $4 \times 4$ antidiagonal  matrix of the form
\begin{equation}
\hat{U}= 
\left(
\begin{array}{llll}
0&0&0&U\\
0&0&-U&0\\
0&-U&0&0\\
U&0&0&0
\end{array}
\right) \;.
\end{equation}
In the above expressions, the  
bare susceptibility $\chi^{(0)}_{\sigma_1 \sigma_2 \sigma_3 \sigma_4} ({\bf q},i\omega_l)$ 
is defined as the convolution of two Green's functions
\begin{subequations} \label{def_chi_zero}
\begin{eqnarray}
&&
\chi^{(0)}_{\sigma_1 \sigma_2 \sigma_3 \sigma_4} ({\bf q},i\omega_l)
=
\nonumber\\
&& \qquad
\sum_{ {\bf k}, i\nu_n} G^{(0)}_{\sigma_1 \sigma_2}  ({\bf k},i\nu_n) 
G^{(0)}_{\sigma_3 \sigma_4}  ( {\bf k}+ {\bf q},i\nu_n+i\omega_l),  \qquad
\label{chi0}
\end{eqnarray}
\noindent with  
\begin{equation}
G^{(0)}_{\sigma_1 \sigma_2} ( {\bf k} , i \nu_n ) = 
\left(\left[ i \nu_n  \sigma_0 - \hat{h} ({\bf k} ) \right]^{-1} \right)_{\sigma_1 \sigma_2}
\label{2x2Green}
\end{equation}
\end{subequations}
the  $2 \times 2$ bare electronic Green's function.
Here,  $\omega_l=2n \pi / \beta$ is the bosonic 
Matsubara frequency, while  $\nu_n = (2n+1) \pi / \beta$ is the fermionic Matsubara frequency,
with $\beta$ the inverse temperature.  

\vspace{0.5cm}

 \section{Simplified expressions for the longitudinal bare susceptibility}
 \label{appendix_B}

In this appendix, we derive simplified expressions for the longitudinal bare spin susceptibility $\chi^{(0)}_{\textrm{long}}$.
First we show that $\chi^{(0)}_{\textrm{long}}$ can be split into 
intra- and interband contributions. For that purpose we note that the four components of the bare Green's function, Eq.~\eqref{2x2Green}, 
can be written as
\begin{eqnarray} \label{Guu1}
G^{(0)}_{\uparrow \uparrow} ( {\bf k} , i\nu_l) 
&=&
G^{(0)}_{\downarrow \downarrow} ( {\bf k} , i\nu_l) 
=
\frac{1/2}{i\nu_l-E_{\bf k}^{1}} + \frac{1/2}{i\nu_l-E_{\bf k}^{2}},
\end{eqnarray}

\noindent and 

\begin{eqnarray} \label{Guu2}
G^{(0)}_{\uparrow \downarrow} ( {\bf k} , i\nu_l) 
&=&
G^{*{(0)}}_{\uparrow \downarrow} ( {\bf k} , i\nu_l)
=
\frac{\hat{{V}}_{\bf k}/2}{i\nu_l-E_{\bf k}^{2}}
-
\frac{\hat{{V}}_{\bf k}/2}{i\nu_l-E_{\bf k}^{1}}, 
\end{eqnarray}
where ${\hat{V}}_{\bf k}=V_{\bf k}/|{V}_{\bf k}|$, with   $V_{\bf k}={\bf g}_{\bf k} \cdot (1,i)$.  
Combining this with Eq.~\eqref{def_chi_zero}, we find that $\chi^{(0)}_{\textrm{long}}$
can be decomposed into an intraband and an interband  part, 
i.e., $\chi^{(0)}_{\textrm{long}} = \chi^{(0), \textrm{intra}}_{\textrm{long}} + \chi^{(0), \textrm{inter}}_{\textrm{long}}$
with 
\begin{widetext}
\begin{equation} \label{myEq_B3}
\chi^{(0), \textrm{intra}}_{\textrm{long}} ({\bf q},i\omega_l)=
\sum_{ {\bf k}} 
\left[\frac{f(E_{\bf k+q}^{1})-f(E_{\bf k}^{1})}{E_{\bf k+q}^{1}-E_{\bf k}^{1}-i\omega_l}+
\frac{f(E_{\bf k+q}^{2})-f(E_{\bf k}^{2})}{E_{\bf k+q}^{2}-E_{\bf k}^{2}-i\omega_l}\right]
\frac{1}{2}(1-{\hat{\bf g}}_{\bf k}\cdot {\hat{\bf g}}_{\bf k+q}),
\end{equation} 
and 
\begin{equation}
\chi^{(0), \textrm{inter}}_{\textrm{long}} ({\bf q},i\omega_l)=
\sum_{ {\bf k}} \left[\frac{f(E_{\bf k+q}^{1})-f(E_{\bf k}^{2})}{E_{\bf k+q}^{1}-E_{\bf k}^{2}-i\omega_l}+
\frac{f(E_{\bf k+q}^{2})-f(E_{\bf k}^{1})}{E_{\bf k+q}^{2}-E_{\bf k}^{1}-i\omega_l}\right]
\frac{1}{2}(1+{\hat{\bf g}}_{\bf k} \cdot {\hat{\bf g}}_{\bf k+q}),
\end{equation}
respectively, where 
$\hat{\bf g}_{\bf k}  = {\bf g}_{\bf k} / | {\bf g}_{\bf k}  |$ , 
$f(z) = ( e^{\beta z} +1 )^{-1}$ is the Fermi distribution function, and $\beta$ is the inverse temperature.

For AFM fluctuations with modulation vector ${\bf q} = {\bf Q} ( 1 + \delta )$ close to $ {\bf Q} = ( \pi, \pi )$, with $\delta \ll 1$, we find that
${\hat{\bf g}}_{\bf k}\cdot {\hat{\bf g}}_{{\bf k} + {\bf Q} ( 1 + \delta )   } = -1 +  \mathcal{O}[t', \delta]$.
Hence, the AFM fluctuations originate from the intraband term, while the interband term gives only a contribution of order $\mathcal{O}[t', \delta]$.
{Thus, we have 
\begin{equation} \label{LQ}
\chi^{(0)}_{\textrm{long}} ({\bf Q},i\omega_l)
\simeq
\chi^{(0),\textrm{intra}}_{\textrm{long}} ({\bf Q},i\omega_l)
=
\sum_{ {\bf k}} 
\left[
\frac{f(E_{\bf k+Q}^{1})-f(E_{\bf k}^{1})}{E_{\bf k+Q}^{1}-E_{\bf k}^{1}-i\omega_l}
+
\frac{f(E_{\bf k+Q}^{2})-f(E_{\bf k}^{2})}{E_{\bf k+Q}^{2}-E_{\bf k}^{2}-i\omega_l}
\right].
\end{equation}

On the other hand, for FM fluctuations with ${\bf q} =  ( \delta, \delta)$ close to ${\bf q}={\bf 0}$, we have  
${\hat{\bf g}}_{\bf k}\cdot {\hat{\bf g}}_{{\bf k} + ( \delta, \delta )   } 
= +1 + \mathcal{O}[\delta^2]$.
Therefore, the FM fluctuations stem from the interband term. 
That is, we find
\begin{equation}   \label{myEq_B6}
\chi^{(0)}_{\textrm{long}} ({\bf 0},i\omega_l) 
=
\chi^{(0), \textrm{inter}}_{\textrm{long}} ({\bf 0},i\omega_l)
=
 \sum_{ {\bf k}} \left[f(E_{\bf k}^{1})-f(E_{\bf k}^{2}) \right]   \left[ \frac{4|{g}_{\bf k}|}{4|{g}_{\bf k}|^2+\omega_l^2}\right] .
\end{equation}
See the main text for a discussion of Eqs.~\eqref{LQ} and~\eqref{myEq_B6}. }

\end{widetext}

\bibliographystyle{apsrev}
\bibliography{rashba-hubbard}

\end{document}